\documentclass[sigconf,natbib=true,anonymous=false,review=false]{acmart}
\usepackage[ruled,linesnumbered]{algorithm2e}
\usepackage{amsmath}
\usepackage{bbm}
\setcopyright{rightsretained}
\acmConference[SIGIR eCom'23]{ACM SIGIR Workshop on eCommerce}{July 27, 2023}{ Taipei, Taiwan}
\acmYear{2023}
\copyrightyear{2023}
\makeatletter
\renewcommand\@formatdoi[1]{\ignorespaces}
\makeatother
\acmISBN{}

\begin{document}

%%\title{XWalk: Candidate Retrieval for Large-Scale Search using Implicit Feedback}
\title{XWalk: Random Walk Based Candidate Retrieval for Product Search}

\author{Jon Eskreis-Winkler}
\email{jeskreiswinkler@etsy.com}
\affiliation{%
  \institution{Etsy, Inc.}
  \country{USA}
}
\author{Yubin Kim}
\email{ykim@etsy.com}
\affiliation{%
  \institution{Etsy, Inc.}
  \country{USA}
}
\author{Andrew Stanton}
\email{astanton@etsy.com}
\affiliation{%
  \institution{Etsy, Inc.}
  \country{USA}
}


\begin{abstract}

%% Old abstract 
% Large-scale search engines are often designed as multi-tiered systems; the candidate retrieval layer efficiently generates a small subset of potentially relevant documents from a corpus many orders of magnitude larger in size, prioritizing recall and latency. 
% %Recent neural retrieval approaches exhibit the ability to infer semantic meaning and extrapolate from user behavioral data, but can underperform simpler methods for common queries, and suffer from latency limitations. 
% In this paper, we propose XWalk, a random walk-based graph approach to candidate retrieval for search in realistic large-scale product search settings with implicit feedback. 
% We demonstrate that XWalk is fast and effective. Our experiments demonstrate that when candidates from XWalk are combined with candidates from a BM25-based inverted index, it substantially improves overall search accuracy while being much quicker to train and inference than more complex neural approaches. We further show improvements compared to state of the art neural models, improving Recall@1000 by 13\% for top queries.  We explore efficiency implications of the architecture, yielding latency improvements of over 76\% compared to existing methods. Finally, we validate the benefits both offline and with online A/B tests.

In e-commerce, head queries account for the vast majority of gross merchandise sales and improvements to head queries are highly impactful to the business.
While most supervised approaches to search perform better in head queries vs. tail queries, we propose a method that further improves head query performance dramatically. We propose XWalk, a random-walk based graph approach to candidate retrieval for product search that borrows from recommendation system techniques. XWalk is highly efficient to train and inference in a large-scale high traffic e-commerce setting, and shows substantial improvements in head query performance over state-of-the-art neural retreivers. Ensembling XWalk with a neural and/or lexical retriever combines the best of both worlds and the resulting retrieval system outperforms all other methods in both offline relevance-based evaluation and in online A/B tests.

\end{abstract}

\maketitle
\section{Introduction}

Modern large-scale search systems are tiered~\cite{Wang2011SIGIR} with at least two layers. The \emph{candidate retrieval} layer generates a small subset of potentially relevant documents from a corpus many orders of magnitude larger in size, while emphasizing efficiency and recall. The \emph{re-ranking} layer uses more computationally expensive methods to re-rank the candidates generated by the retrieval stage to produce a high-precision final result list. Better recall in candidate retrieval leads to better overall accuracy. In this paper, we focus on improve search through improving recall in the candidate retrieval layer.

Most evaluations for search systems use an evaluation query set in which every query is assumed to be equally important and has equal impact on the accuracy metric. However, in reality, query frequency distributions are exponential~\cite{yates_tiberi_2007}. Consequently, in e-commerce, head queries account for the vast majority of gross merchandise sales and head query performance is far more impactful to business metrics than torso or tail performance. 
State-of-the-art supervised neural dense retrievers \cite{lee_latent_2019,karpukhin_dense_2020,xiong_approximate_2021,chen_out--domain_2022,wang_bert-based_2021} typically perform better in head queries than tail, due to the higher availability of training data in the head region. However, we show that further substantial improvements to head query performance are possible. We borrow ideas from the recommendation systems community and propose XWalk, a graph-based approach to candidate retrieval.

Historically, graph-based approaches in search were used to create features (e.g. PageRank, click graphs~\cite{jiang_learning_2016,zhang_neural_2019}) for the re-ranker layer, but have not been used directly for retrieval.
Recently, graph neural networks (GNNs) have achieved state of the art performance in recommendation and are being adapted for search \cite{li_learning_2020,zamani_learning_2020,xia_searchgcn_2021,zhao_joint_2022}. However, large-scale GNNs are complex and slow to train.  

The recommendation systems have long used implicit interaction graphs to directly generate recommendations. Commonly, users and product listings are represented as nodes in a graph and edges represent a logged interaction between a user and product listing (e.g. the user purchasing the listing). 
Random walks in graphs is a powerful technique used to generate recommendations from interaction graphs \cite{park_comparative_2017,christoffel_blockbusters_2015,eksombatchai_pixie_2018,paudel_updatable_2016}.
Random walk based approaches are frequently used in large, real-time recommendation systems due to their effectiveness and efficiency \cite{paudel_updatable_2016,eksombatchai_pixie_2018}. In addition, when using implicit feedback (e.g. logged interaction data such as user clicks) Park et al. \cite{park_comparative_2017} showed that random walk based approaches can perform better than matrix factorization approaches. 

XWalk uses a random walk based approach to perform candidate retrieval for product search. In XWalk, we cast search as a query-to-listing recommendation problem (as opposed to user-to-listing), that is, we transform our query log into a implicit interaction graph between queries and product listings, and perform candidate retrieval by ``recommending'' listings to queries. Our approach trains using a fraction of the time and resources used by neural dense retrievers and GNNs, and is highly efficient in inference -- XWalk scales to real-time search over graphs of billions of nodes and tens of billions of edges. XWalk also excels in head queries, where implicit feedback signals are plentiful. 

While XWalk on its own suffers in tail and novel queries, we show that when results from XWalk are ensembled with a typical retriever that uses text similarity, even one as basic as plain BM25, it substantially improves overall candidate retrieval accuracy compared to strong neural dense retrieval and hybrid retrieval baselines, especially over the head query region, which is responsible for the overwhelming majority of sales in e-commerce. Furthermore, we show that XWalk is complementary to \emph{both} dense retrieval and BM25, and demonstrate the strength of ensembling all three approaches.

To summarize, our novel contributions are: a) showing that XWalk substantially improves performance in the head query region, which accounts for the overwhelming majority of sales in e-commerce;
b) presenting an efficient random walk inference algorithm that can effectively serve queries at scale;
c) showing that XWalk is complementary to other common retrieval methods and showing the strength of a simple ensemble approach that combines XWalk, BM25, and dense retrieval.

\section{Method}

% Etsy inventory is unlike other inventory in e-commerce; artisans craft unique items that defy normal taxonomic categorization, are often one of a kind, and typically align along a individual niches.  When taking step back and looking at the marketplace as whole, we observe Etsy is in practice a loosely connected set of thousands individual marketplaces with sellers dedicated to their individual areas of specialization.  In particular, queries are highly conditioned on the specific area of interest - that is, common search terms shift radically depending on the particular niche of choice (TODO: Example Needed).  This ambiguity is especially challenging for traditional term matching techniques and require a rethinking of the problem.

%To capture community sentiment scalably, 
We take inspiration from the recommendation space and recast the search problem as a query-to-product listing recommendation problem using implicit feedback: predict the best $k$ product listings $L_{q_i}$ to ``recommend'' to a query $q_i$, by learning from implicit user feedback, i.e. a query log.
From the query log, we construct an undirected, weighted bipartite graph $G = (Q,L,E,W)$ where $Q$ are nodes representing queries, $L$ are nodes representing product listings, $E$ are edges $E = \{e_{i,j} = (q_i, l_j) \mid q_i \in Q \land l_j \in L\}$, and $W$ are edge weights.

% One challenge of encoding raw user feedback into an unweighted graphs is capturing popularity.  Due to the sparse community structure in the Etsy marketplace, edge weighting becomes necessary; while graphs often make the assumption that node degree is a reasonable proxy for popularity(TODO: Add reference to Fortunato paper), in this framing node degree also capture generality - that is, the greater the cardinality of queries associated with a listing, the more likely it's satisfying a discovery oriented query (e.g. "gift").  To account for this, we weight the edges $\it{w}$ to capture Query to Listing popularity.

\subsection{Graph Construction (Offline Training)}

Given a query log which records for each query $q_i$ the set of listings $L_i^{click},L_i^{cart},L_i^{purchase} $ that the user clicked on, added to their shopping cart, and purchased, respectively, we construct our graph through the following process:

\begin{enumerate}
    \item For each unique (by text string) query in the query log $\hat{q}_i$, add $\hat{q}_i$ to $Q$.
    \item For each unique (by listing ID) listing in the query log $l_j$, add $l_j$ to $L$.
    \item Collate the query log by query-listing pairs $(\hat{q}_i, l_j)$, counting the number of occurrences of $click_{i,j}$, $cart_{i,j}$, and $purchase_{i,j}$ interactions for each unique $(\hat{q}_i, l_j)$ pair. 
    \item For each $(\hat{q}_i, l_j)$, add $e_{i,j}$ to $E$ and its weight $w_{i,j}$ to $W$, where $w_{i,j}$ is calculated Equation \ref{eq:weighted_edge}. 
\end{enumerate}

Intuitively, edge weights represent the popularity or trustworthiness of the edge, i.e. if many different users bought listing $l_i$ from query $q_j$, $w_{i,j}$ will be higher because we are more confident in the relationship represented by the edge. To weight edges, we use a simple linear combination:
\begin{equation} \label{eq:weighted_edge}
    w_{i,j} = C_1\cdot\it{|click_{i,j}|} + C_2\cdot{\it{|cart_{i,j}|}} + C_3\cdot{\it{|purchase_{i,j}|}}
\end{equation}

In practice, the best coefficients are $C_1 < C_2 < C_3$, as the goal is to bias walks toward listings which convert well for a given query.  

\subsubsection{Graph representation for efficient inference}
    %\item For each node ($\{Q, L\}$), convert weights into CDF format for Inverse Transform Sampling.
    %\item Convert the graph into Compressed Sparse Row format.  Sort the nodes in descending order of node degree.
XWalk is designed for sparse graphs scaling up to billions of nodes and tens of billions of edges. The costliest part of random walk graph inference is sampling edges to walk, especially from high degree nodes. For efficient inference, we choose our graph representation carefully.

We store edge weights as cumulative distribution functions in order to use Inverse Transform Sampling,
%(citation to Knight paper) 
which allows sampling in $O(log(N))$ time. Note, we choose this approach over the alias method,
%(citation to smola paper), 
which allows for constant time sampling, due to the doubling of memory needed for the transform.  As XWalk's space complexity is dominated by edges and corresponding weights, we develop other methods for efficient sampling (Section \ref{weighted-sampling}).

To transform edge weights in to CDF format, for each node $n$, we sort its adjacent edges $E_{n,*}$ in decreasing order of their weights $W_{n,*}$, such that $w_{n,i} > w_{n,i+1}$.  We then compute the cumulative distribution of all weights:

\begin{equation} \label{eq:cdf}
    CDF_{n,i} = \frac{\sum_{j=0}^{i}{w_{n,j}}}{\sum_{i=0}^{|E_{n,*}|}w_{n,i}}
\end{equation}

To sample an edge from $E_{n,*}$, we randomly sample $p \sim Uniform(0,1)$ and find the corresponding edge through binary search. This formulations provides us a few valuable advantages:
\begin{enumerate}
    \item Weighted sampling is $O(log(|E_{n,*}|)$.  Given some nodes have degrees in the millions, logarithmic growth is critical for performance.
%    \item Due to sorting by decreasing weight, we can retrieve top K neighbors in constant time.
    \item Normalizing the CDF to 1 allows us to reconstruct the the transition probability for outbound edges.  This is key for the Metropolis-Hastings sampling strategy (Section \ref{weighted-sampling}).
    \item Better cache coherence as the bulk of the weights are located near the front of the distribution.
\end{enumerate}

Finally, we convert the graph into Compressed Sparse Row format, guaranteeing a $O(1)$ lookup cost for edges. % and sort the nodes in descending order of node degree.

Note that all of the above graph construction steps are simple ETL (extract, transform, load) operations with no expensive parameter training steps. Compared to neural dense retrievers, ``training'' an XWalk graph model takes only a fraction of the cost and time.

\subsection{Graph Inference (At Query Time)}

Inferencing a graph with random walks is challenging to do efficiently. Despite the $O(1)$ edge lookup guarantee of the Compressed Sparse Row format used in graph construction, a naive walk approach that uses depth first search and binary search node lookups create random memory access patterns which result in high rates of costly cache misses~\cite{yang_random_2021}. We present an approach for XWalk that scales to graphs of billions of nodes and tens of billions of edges.

At query time, XWalk retrieves relevant listings for a query $q_i$ by sampling nodes in $G$ using $k$-hop fixed paths \cite{christoffel_blockbusters_2015,eksombatchai_pixie_2018} with node $q_i$ as the starting point. When $k$ is an odd number, the last node in a $k$-hop path will always be a listing node ($L$) due to the bipartite nature of $G$. XWalk returns listings ranked by the frequency of which they were sampled.

\label{weighted-sampling}

To reduce costly random memory access patterns, we use a breadth first search instead of depth first search for our random walks. We also improve upon the Inverse Transform Sampling strategy by using the Metropolis-Hastings algorithm (a Markov chain Monte Carlo method) in most places. 
 Given the sorted CDF format of edge weights (Eq. \ref{eq:cdf}), we can reconstruct the original edge transition probabilities: $P(n_j|E_{n_i,*}) = w_{n_i,j} - w_{n_i,j-1}$. As Metropolis-Hastings requires a symmetric distribution, we take the absolute value of the proposal index for each edge and sample from the Normal distribution.  Ablation testing indicated XWalk is not sensitive to the variance for the proposal distribution, $\sigma^2$. We set $\sigma^2 = 0.2$.
 
Metropolis-Hastings improves the cost of $c$ edge samples to $O(\log(|E_{n,*}|)) + c$ compared to $c*O(\log(|E_{n,*}|))$ of Inverse Transform Sampling. In cases where $c$ is large (e.g. the initial query node), the computational improvements are substantial.  A known limitation of MCMC methods is the auto-correlation of samples, usually requiring a mix time prior to sampling.  Therefore, for our first sample, we use Inverse Transform Sampling to get an unbiased starting point and use Metropolis-Hastings for subsequent samples. In preliminary testing we found no reduction in model accuracy for this implementation compared to using only Inverse Transform Sampling while seeing the expected substantial latency benefits. 
 
Our overall random walk strategy is presented in Algorithm \ref{alg:xwalk-bfs}.
\vspace{-1 em}

%Details of Metropolis-Hastings is presented in Algorithm \ref{alg:metropolis}. 

%\begin{algorithm}[h]
%\textbf{Global variables: } 
%Variance $\sigma^2$,
%Dictionary of nodes to counts $Counter$ \\
%\textbf{Input: } 
%Starting node $n$,
%Number of walks $c$,
%Walk-length $k$,
%Edges $E$,
%Weights $W$,
%Multiplier $m$ (default 1) \\
%$p \sim Uniform(0, 1)$ \\
%$i = BinarySearch(E_{n,*}, W_{n,*}, p)$ \\
% \tcc{the $i$'th node of ordered neighbors of $n$}
%$Counter[node(E_{n,i})] += m $ \\
%\For{step = \{2, .., c\}} {
%    $j = Metropolis(i, E_{n_i,*}, W_{n_i,*}, \sigma^2)$ \\
%    \tcc{the $j$'th node of ordered neighbors of $n_i$}
%    $Counter[node(E_{n_i,j})] += m$ \\
%    $i = j$
%}
%\uIf{\text{k > 0}} {
%    $counts = \emptyset$ \\
%    \For{\{n_i, count\} \in Counter} {
%        $counts = counts \cup XWalkBFSSampler(n_i, c, k-1, E, W, count)$
%    }
%    \Return $counts$ 
%}\Else{
%    $ Nodes = Sort(node \in Counter) \text{ in non-increasing order } \{Counter[n_1] \geq Counter[n_2] \ldots \geq  Counter[n_{\lambda}]\} $ \\
%    \Return Nodes \\
%}
%\caption{XWalkBFSSampler}
%\label{alg:xwalk-bfs}
%\end{algorithm}

\begin{algorithm}[h]
\textbf{Global variables: } 
Var of Normal distribution $\sigma^2$,
Dictionary of nodes to counts $Counter$ \\
\textbf{Input: } 
Starting node $n$,
Number of walks $c$,
Walk-length $k$,
Edges $E$,
Weights $W$,
Multiplier $m$ (default 1) \\ 
$p \sim Uniform(0, 1)$ \\
$i = BinarySearch(E_{n,*}, W_{n,*}, p)$ \\
 \tcc{the $i$'th node of ordered neighbors of $n$}
$Counter[node(E_{n,i})] += m $ \\
\For{step = \{2, .., c\}} {
    $j = Metropolis(i, E_{n_i,*}, W_{n_i,*}, \sigma^2)$ \\
    \tcc{the $j$'th node of ordered neighbors of $n_i$}
    $Counter[node(E_{n_i,j})] += 1$ \\
    $i = j$
}
\uIf{\text{k > 0}} {
    $counts = \emptyset$ \\
    \For{$\{n_i, count\} \in Counter$} {
        $counts = counts \cup XWalkBFSSampler(n_i, count, k-1, E, W)$
    }
    \Return $counts$
}  \Else {
    $ Nodes = Sort(node \in Counter) \text{ in non-increasing order } \{Counter[n_1] \geq Counter[n_2] \ldots \geq  Counter[n_{|Counter|}]\} $ \\
    \Return Nodes \\
}
\caption{XWalkBFSSampler}
\label{alg:xwalk-bfs}
\end{algorithm}

% \begin{algorithm}[h]
% \textbf{Input: } 
% Index $i$,
% Edges $E_{n_i,*}$,
% Weights $W_{n_i,*}$,
% Variance $\sigma^2$ \\
% $p \sim Normal(0, \sigma^2)$ \\
% $proposal = \lfloor |i + p \cdot |E_{n_i,*}|| \rfloor$ \\
% \uIf {$proposal > |E_{n_i,*}|$} {
%     \Return $i$
% }\Else {
%     $acceptance = \frac{w_{n_i, proposal}}{w_{n_i, i}}$ \\
%     \uIf{a \sim Uniform(0,1); a < acceptance} {
%         \Return proposal
%     }\Else {
%         \Return i
%     }
% }
% \caption{Metropolis}
% \label{alg:metropolis}
% \end{algorithm}

\vspace{-2 em}
\subsection{Extending the Graph}
\label{section:shoptags}
Our e-commerce platform is a two-sided marketplace and our inventory comes from independent sellers. Thus, listings are naturally grouped by shops. In addition, sellers may add tags to their listings to better describe them (e.g. ``christmas'', ``gift'', etc.). 

For the sake of notation simplicity, we described the graph construction and inference above assuming our graph only contains two types of nodes, $Q$ and $L$. However, in practice, we extend the graph by adding shop nodes ($S$) and tag nodes ($T$) to the graph; this allows us to retrieve listings without implicit user feedback (e.g. the cold start problem) and further increase connectivity of the graph. Note that $G$ remains bipartite: $\{Q, S, T\}$ is a separate partition from $L$ and thus the algorithms described in this section can be used unchanged. The weights of edges between shops/tags and listings are set to 1. $w_{q_i,s_j} = w_{q_i,t_j} = 1$.

\section{Experiments}

For our experiments, we sought to closely emulate a real-world e-commerce setting, where the main source of training data is implicit user feedback from query logs, and models are evaluated under a realistic query popularity distribution. Unfortunately, most public search datasets do not reflect a realistic query distribution and rarely have implicit user feedback as training data. While recommendation system datasets have implicit user feedback, they do not have a text query that is usable by BM25 or dense retrieval, which retrieve based on query to listing text similarity. Therefore, we curated a training and evaluation dataset from our e-commerce platform.

\subsection{Dataset Creation}

For training data, we collected 365 days of implicit feedback data, comprising of records of queries and the product listings that were clicked, added to cart, or purchased from a given query. 
Queries are represented by their query text. Listings are represented by their unique ID and the title of the product. In addition, as mentioned in Section~\ref{section:shoptags}, listings are associated with seller-provided tags, and each listing belongs to exactly one seller's shop.

Over the time period used for this experiment there were 137,824,871 unique listings, 147,174,817 unique queries, 62,803,463 unique tag, and 3,018,713 unique shops. 
There were a total of 1,349,734,328 query-listing interactions recorded, where 3.46\% were purchases, 6.19\% were cart adds and 90.3\% were clicks. Altogether, there were 1,395,759,140 edges. 
Example records are found in Table \ref{sample_record}.

\begin{table*}[!h]
\begin{tabular}{ c c  c  c  c c } \hline 
   query  &  listing ID & listing title & interaction type & shop & tags  \\ \hline \hline
   wedding dress & l12 & beautiful bridal wedding gown & click & s00 & white, gown \\
   wedding gown & l12 & custom embroidered wedding dress & purchase & s00 & white, gown \\
   wedding dress & l34 & ethereal champagne dress with chiffon skirt & click & s11 & fancy, dress, chiffon \\
   \hline 
\end{tabular}
\caption{Hand-created example of implicit feedback training data.}
\label{sample_record}
\vspace{-2em}
\end{table*}

% Each interaction is a query, a listing, and an associated interaction.  All together this comprises a bipartite graph with four types of nodes: query, listing, shop and tag. Edges exist between listing node and a query node if an interaction occurred in the lookback period; edges exist between a listing node and a shop node if the listing is sold by the shop; edges exist between a listing node and a tag node if the listing’s seller described their listing using that tag. 

\begin{figure}[h]
\centering
\includegraphics[width=0.35\textwidth]{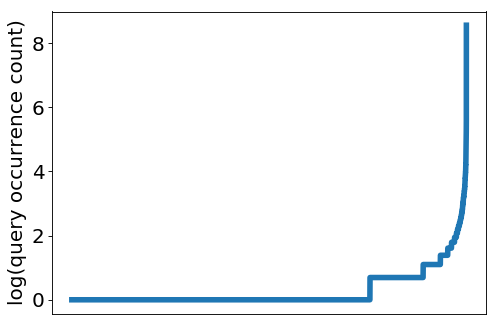} \\
\caption{Distribution of evaluation queries' log frequency, ordered from least frequent to most frequent.}
\label{querycount}
\vspace{-2em}
\end{figure}

Evaluation data was curated to be a representative query distribution, sampled from a single day immediately following the last day of the training data window. We randomly sampled 11,521 queries that resulted in at least one purchase. As the sample is intended to be reflective of the true query popularity distribution, we did not de-duplicate the query set. Figure~\ref{querycount} shows the distribution of the query frequency in the evaluation set.

For each query, the listings that were purchased from that query are considered the relevant document. 82.3\% of queries had only one purchase, 12.0\% had two purchases and 5.6\% had more than two purchases. For each one of these queries, we assigned them to a head/torso/tail frequency bin based on how frequently they occurred in the previous 365 day period. The bins were created such that the total counts of requests are roughly equal among those bins. Of the evaluation queries, 31.0\% were in the head bin, 47.9\% were in the torso bin, and 43.9\% were in the tail bin.

\subsection{Experiment set up}

We compare XWalk against two other methods of candidate retrieval. First is lexical retrieval using BM25 scoring (BM25). We use Pyserini \cite{Lin_etal_SIGIR2021_Pyserini} to build a Lucene index based on listing titles in our dataset and then retrieve candidates using BM25 rankings using bag-of-words representations. We used the default analyzer and default BM25 parameters (k1=1.2, b=0.75).

% The details regarding the neural baseline (NIR) were not explained clearly and in full detail. What was the technique, how much was the training data used? for how many epochs was it trained for? Without knowing these details, it's very hard to believe the claim that the XWalk technique performs better than NIR.

The second baseline is a state of the art neural dense retrieval system \cite{nigam2019} trained on search traffic for candidate retrieval (NIR). NIR uses a smaller time window of training data (30 days) due to the time and expense of training on larger data sets. NIR is a Transformer-based, two tower model that uses a multi-part hinge loss to distinguish between interactions that involve a purchase, cart add, favorite, click, or nothing. The model was trained over one epoch. %for YYY hours. 
It was designed for better semantic matching between queries and listings by incorporating title, query as well as additional features such as tags, and listing taxonomy.

In addition to the above, we also compare results against hybrid systems of NIR+BM25 \cite{chen_out--domain_2022}. 
To ensemble the results from each retrieval engine, we use Reciprocal Rank Fusion, a simple but effective fusion technique \cite{chen_out--domain_2022}.
%We emphasize now that our argument in this paper is to demonstrate that the marriage of pure lexical search with a non-semantic interaction based retrieval outperforms each individually, so we explore how the combination of our fast graph-based retrieval service, together with the benchmark lexical model produces high quality results. 
Higher recall in candidate retrieval result in higher overall search accuracy~\cite{chen_out--domain_2022}.
As we are focusing on the first-pass candidate retrieval stage of search, 
we use recall and mean average precision (MAP) at 100, and 1000 to measure the quality of candidates retrieved. %Note that our dataset defines previously purchased listings as ``relevant'', thus recall and MAP measure recall of previously purchased listings.

%In showing how these forms of recall can be combined, candidate sets are generated by stochastically combining the candidate sets from XWalk and from BM25. For demonstrative purposes, we computationally flip a biased coin (with heads probability equal to $\alpha$) to determine whether to draw Xwalk (heads) or BM25 (tails) for each position, with higher values of $\alpha$ leading to more weight on XWalk candidates and lower values of $\alpha$ leading to more weight on BM25 candidates.

\section{Analysis}

As shown in Table~\ref{tab:main}, when compared independently against other methods, XWalk out-performs other methods in most metrics despite the fact that it is unable to return results for novel queries, due to its strength in the head query bin. When combined with BM25, it outperforms in every metric, both NIR and the hybrid NIR+BM25. Finally, the ensemble of all three methods (XWalk+BM25+NIR) substantially outperforms all other configurations.

%The reason for the above is evident when examining queries segmented by their popularity bins in Table~\ref{bins_recall}.
We see in Table~\ref{tab:main} that BM25 is significantly weaker in performance compared to NIR and XWalk and does not always improve the overall results of NIR and XWalk, especially for MAP. While BM25 can improve recall by adding listings that were not retrieved by NIR and XWalk, its poor ranking drags down MAP in the hybrid systems. For the most popular short queries, BM25 is not able to distinguish between the many listings with titles that token match similarly to the query. Whereas methods like XWalk and NIR are able to provide a more reliable ranking of the highly purchaseable listings based on training data. 

However, in Table~\ref{bins_recall} we see that XWalk is complementary to BM25; XWalk is stronger in the head and torso bins while BM25 outperforms XWalk in the tail bin. This is due to the fact that XWalk suffers from cold start problems: it performs best with many prior examples and is unable to handle novel queries. BM25, as a lexical matching system, is more able to handle novel queries. Furthermore, XWalk+BM25 is still yet complementary with NIR. The semantic matching of dense retrieval excels in the tail, where queries are typically longer. When all three systems are ensembled, it is the highest performing across all query bins. XWalk's success in the head query bin is particularly notable -- in an e-commerce setting, the head query bin is responsible for a large majority of merchandise sales.

%Despite that XWalk is not able to 
%
%Unsurprisingly, using only a lexical model misses much of the meaningful information contained in historic user interactions. As a standalone model, lexical matching trails far behind the random walk based approach. Looking deeper, however, we find that combining XWalk with lexical matching exceeds XWalk’s performance on it’s own - specifically in the areas where it performs worst. Whereas XWalk has highest recall and MAP for the most popular queries (top.01 and top.1), for BM25 the highest metrics are for the least common queries (tail and no bin). This is evident in figures \ref{recall100} - \ref{map1000} where $\alpha=0$ is pure lexical matching and $\alpha=1$ is pure XWalk.

\begin{table}[!h]
\begin{tabular}{ l  c  c   c  c  }
 \hline
 & r@100 & r@1000 & M@100 & M@1000 \\
 \hline
 \hline
BM25 & 0.192 & 0.394  & 0.034 & 0.035 \\
NIR & 0.547 & 0.740  & 0.107  & 0.109 \\
XWalk & 0.600 & 0.723 & 0.153 & 0.154 \\
NIR+BM25 & 0.497 & 0.780  & 0.097 & 0.100 \\
XWalk+BM25 & 0.599 & 0.829 & 0.129 & 0.132  \\
% XWalk+NIR &  0.621 & 0.902 & 0.193 &  0.197 \\
XWalk+BM25+NIR & \textbf{0.701} & \textbf{0.915} & \textbf{0.194} & \textbf{0.198}  \\
 \hline
\end{tabular}
\caption{recall (r@100, r@1000) and MAP (M@100, M@1000) for retrieval models and combinations.}
\label{tab:main}
\end{table}
\vspace{-3em}
\begin{table}[!h]
\begin{tabular}{ l  c  c  c  }
\hline 
 & tail & torso & head \\
 \hline
 \hline
BM25 & 0.471 & 0.420 & 0.299 \\
NIR & 0.738 & 0.728 & 0.759 \\
XWalk & 0.260 & 0.813 & 0.899 \\
NIR+BM25 & 0.804 & 0.779 & 0.762 \\
XWalk+BM25 & 0.595 & 0.875 & 0.914 \\
% XWalk+NIR & 0.794 & 0.923 & 0.941 \\
XWalk+BM25+NIR & \textbf{0.836} & \textbf{0.931} & \textbf{0.942} \\
\hline
\end{tabular}
\caption{Comparison of retrieval models in terms of recall@1000 stratified by query popularity.}
\label{bins_recall}
\end{table}

\vspace{-2em}

% \begin{table}[!h]
% \begin{tabular}{ l  c  c  c  }
% \hline 
%  & tail & torso & head \\
%  \hline
%  \hline
% BM25 & 0.070  & 0.039 & 0.006 \\
% NIR & 0.135  & 0.122 & 0.072 \\
% XWalk & 0.062 & 0.181 & 0.174 \\
% NIR+BM25 & 0.146 & 0.114 & 0.049\\
% XWalk+BM25 & 0.107 & 0.150 & 0.115 \\
% % XWalk+NIR & 0.170 & \textbf{0.224} & \textbf{0.177}\\
% XWalk+BM25+NIR & \textbf{0.175} & \textbf{0.222} & \textbf{0.173} \\
% \hline
% \end{tabular}
% \label{bins_map}
% \caption{Comparison of retrieval models in terms of MAP@1000 stratified by query popularity}
% \end{table}

%The critical region of these plots is for high values of $\alpha$ where we find that we gain almost all the benefits of XWalk with a boost from lexical signals coming from BM25. Of specific note is the difference between high and low frequency queries. While for popular queries it is not possible to differentiate between pure XWalk and high values of $\alpha<1$, there is a sudden shift for the least common queries. This in turn leads to a jump in recall and MAP over all queries when $\alpha \approx 0.95$ versus when $\alpha=1$.

\section{Online Testing}

%In an online A/B experiment, visiting browsers are placed into a control or variant bucket and are given the experience for the entire duration of the online experiment. At the conclusion of the experiment, we compare the conversion rate of browsers (i.e. the percentage of browsers that made a purchase) placed in the control bucket vs. the variant bucket.

We tested XWalk in a live online A/B experiment on a large e-commerce platform. The experiment ran for 23 and 25 days on our mobile and web version of our platform, respectively. Our search system is a two-stage search system, which uses an ensemble of candidate retrievers in the first pass, followed by a second pass re-ranker.  In our A/B experiment, an ensemble of NIR+Solr as the candidate retrieval system was compared against an ensemble of XWalk+NIR+Solr.

We saw a statistically significant and substantial increase in conversion rate for the search system including XWalk in both the web and mobile platforms, $+1.2\%$ on web and $+1.98\%$ on mobile. In addition, in a production setting, we saw that XWalk was our lowest latency retrieval engine. The 99th percentile latency is only 58\% of the NIR engine and 22\% that of our Solr inverted index. 

%We ran several online tests on a large e-Commerce website across all queries to test our hypotheses regarding the contribution of interaction-based retrieval. We observed significantly positive increases in conversion rate, click-through rate, revenue, and many other secondary metrics. We also anecdotally found that search results from XWalk presented some clever examples of non-semantically similar relevant listings. For example, when users searched for a TV show, some results featured actors on the show or spinoff series.
%
%Some engineering effort was expended in developing a service for mixing retrieved candidates from different sources but we have reaped the dividends of this effort as we continue to explore new ways to combine complementary candidate retrieval services. One challenging regime has been learning how to apply XWalk to novel queries. Subsequent work improving our lexical retrieval service coupled with developments in neural IR have mitigated what might otherwise have been an insurmountable block in using a graph-based retrieval service. 
%
%Perhaps most importantly, we observed massive latency gains as projected offline. This is attributed to using a low-level language like Rust\cite{matsakis2014rust} for implementation. Specifically, we found that XWalk returned candidates on average >75\% more quickly than Apache SOLR. 

% \begin{figure}[h]
% \centering
% \includegraphics[width=0.35\textwidth]{recall@100.png} \\
% \caption{recall@100}
% \label{recall100}
% \end{figure}

\section{Conclusion}
Head queries are responsible for the large majority of purchases in e-commerce. We presented XWalk, a novel candidate retrieval engine, which by frames search as a query-to-product recommendation problem, leverages powerful, highly efficient graph methods to substantially improve head query performance in product search. XWalk is also complementary to other common retrieval engines such as BM25 and dense retrieval, and ensembling produces a powerful retrieval engine.

\bibliography{sigir2023-xwalk}

\end{document}